# Whispering-gallery modes promote enhanced optical backflow in a perforated dielectric microsphere

Yury E. Geints,[1] Igor V. Minin,[2] Oleg V. Minin[2,*]

[1]*V.E. Zuev Institute of Atmospheric Optics, 1 Acad. Zuev square, Tomsk, 634021, Russia*
[2]*Tomsk Polytechnic University, Lenina 36, Tomsk, 634050, Russia*

**Optical energy flow inside a dielectric microsphere is usually codirected with the optical wavevector. At the same time, if the optical field in a microsphere is in resonance with one of the high-quality spatial eigenmodes (whispering-gallery modes - WGMs), a region of reverse energy flow emerges in the shadow hemisphere. This area is of considerable practical interest due to increased optical trapping potential. In this Letter, we consider a perforated microsphere with an air-filled single pinhole fabricated along the particle diameter and numerically analyze the peculiarities of WGM excitation in a nanostructured microsphere. A pinhole isolates the energy backflow region of a resonant mode and changes a perforated microsphere into an efficient optical tweezer. For the first time to our knowledge, we reveal the multiple enhancement of backflow intensity in the pinhole at a WGM resonance and discuss the way for its manipulation.**

**Keywords**: mesoscale particle, WGM, pinhole, backflow optical energy

An important property of the spatial regions with extreme optical field concentration in mesowavelength dielectric particles is the appearance of phase singularity points in the form of optical vortices which may results in the formation of a reversed optical energy flow directed toward the propagation of incident optical wave [Bovin1967]. Particularly, a reverse energy flux occurs when optical radiation diffracts through a subwavelength aperture [Schouten2004], by focusing propagation of high-order cylindrical vector beams, in vortex optical fields with large topological charges passing through a complex gradient-index (GRIN) lens [Li2020, Stafeev2020], in a dielectric wavelength- scaled (mesoscale) microparticle exposed by a complex-structured vector beams [Wang2021] etc. Such mesoscale dielectric particles represent a promising family of wavelength-scaled optics possessing an elegant, mostly subwavelength, low-loss focusing of optical radiation [Luk'yanchuk2017, Chen2018]. Worth noting, all above-mentioned cases require specially structured irradiation or specific refractive index structured focusing lenses.

The generation of structured optical beams for various purposes is possible by nanostructuring the shadow surface of mesoscale particles [Wu2015, Zhou2018, Zhou2021]. As shown in [Cao2019], the fabrication of a single nanohole inside a dielectric microparticle allows for strong optical field localization near hole opening with transverse dimensions depending only on the hole diameter and refractive index contrast between hole and particle regardless the illuminating wavelength. This method makes it possible creation a near-field localized beam with minimal waist far beyond the diffraction limit, which opens up new prospects, in particular, for optical manipulations with nanoparticles [Minin2019]. Importantly, the realization of optical field focusing inside a nanohole can be used to separate the regions of forward and backward energy fluxes [Minin2021]. However, increasing the intensity of the backward energy flux by carving a specific nano-relief in a microparticle remains challenging.

So far, the problem for creating an intense optical energy backflow in microparticles was considered in the, so-called, scattering mode when no optical resonances are excited in particle volume. In this work, for the first time to the best of our knowledge, we investigate the physical origin of a resonant reversed energy flow formation in a mesoscale dielectric sphere structured with an open-end pinhole. By numerical experiments we show that the reversal of the optical energy flux occurs inside the hole under the conditions of an electromagnetic resonance - the whispering gallery mode (WGM). Importantly, the energy backflow is several times greater than the forward-directed flux and depends on the nano-hole diameter and the quality of the excited WGM. Worth noting, resonant backflow amplitude significantly exceeds the corresponding value in the nonresonant mode.

Consider a spherical dielectric microsphere with radius $R = 1$ μm and refractive index $n = 2$ (e.g., polycrystalline barium borate) located in air and illuminated by a plane monochromatic circularly polarized optical wave with the wavevector $k_z$ and a visible wavelength. The dimensionless Mie size-parameter of a microparticle, $\rho = 2\pi R/\lambda$, in this case amounts to $\rho \sim 15$ that falls within the mesoscale range [Luk'yanchuk2017]. Inside the microsphere, along its principal diameter in the direction of light incidence a cylindrical pinhole is fabricated with the diameter $d_h \ll \lambda$, which is filled with air.

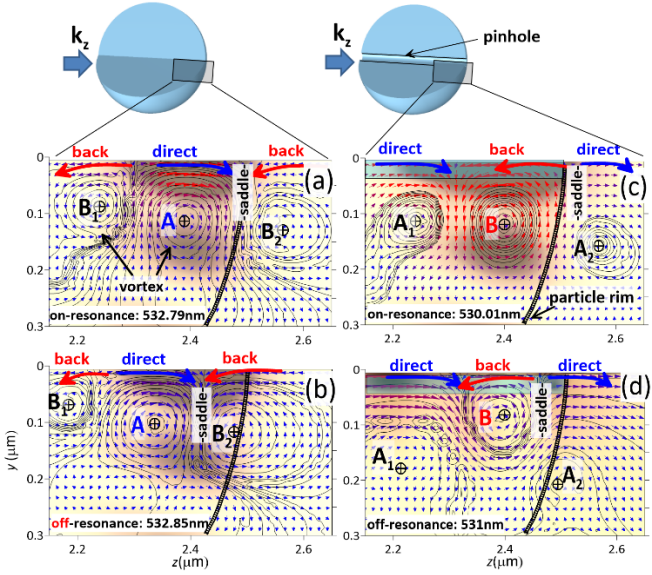

Fig. 1. 2D-distributions of Pointing vector **S**(y, z) near the shadow hemisphere of (a),(b) solid and (c),(d) perforated microparticle. A 60 nm pinhole is depicted by shaded rectangle.

In the simulations, the stationary Helmholtz equation for the electric field vector is solved using the finite element method implemented in the COMSOL Multiphysics package. The spatial resolution of the computational tetrahedral grid is employed with mesh size less than $\lambda/(15n)$ in the regions outside the hole and $d_h/5$ inside it. For solution accuracy improving, all boundary edges between the heterogeneous domains are extra-meshed with multiples of smaller grid step. At domain outer boundaries, wave free propagation conditions are applied by constructing a set of perfectly matched layers (PML).

Consider the structure of optical energy flow in the conditions of a WGM resonance in the microsphere based on the time-averaged Poynting vector field $\mathbf{S} = (c/8\pi)\operatorname{Re}\left[\mathbf{E}\times\mathbf{H}^*\right]$ (**E** and **H** are the electric and magnetic field, respectively $c$ is the speed of light). Figures 1(a)-(d) visualize the optical energy flows in the vicinity of the shadow pole of a solid (left column) and perforated (right column) spherical particle near $TE_{19,1}$ mode resonance. The figures show the two-dimensional (y-z) maps of Poynting vector directions. Note, the arrow size depends on the energy flux strength |**S**| and **S** streamlines are plotted here with solid contours also. As seen, the Poynting vector fields are highly turbulent, so we several characteristic topological phase defects can be distinguished, namely, points of phase singularity around which the optical vortices are organized (marked with bold crosses), and located between the vortices phase field saddle zones, where the longitudinal Poynting component $S_z$ vanishes. Depending on the direction of energy circulation (spin), the phase singularity points are divided into vortices with clockwise circulation (type A) and counterclockwise circulation (type B). In the framework of the problem geometry, the A-vortex (negative vortex [Schouten2004]) forms a direct optical flow along the main diameter of microsphere (at $y = 0$) in the direction of the incident wavevector $\mathbf{k}_z$ (shown by blue bold arrow), while the positive B-vortex, on the contrary, launches a reverse axial energy motion (red bold arrows).

As known, according to the physical origin, the Poynting flux can be divided into a potential (orbital) $\mathbf{S}_O$ and solenoidal (spin) $\mathbf{S}_C$ [Bekshaev2007, Berry2009]: $\mathbf{S} = \mathbf{S}_O + \mathbf{S}_C$. The orbital flux component is directly related to the wave phase gradient $\nabla\varphi$ and is responsible for the appearance of the inverse optical currents, $S_z \equiv \mathbf{e}_z \cdot \mathbf{S}_O = I/k\left(\partial\varphi/\partial z\right) < 0$, where $I = c/8\pi|\mathbf{E}|^2$ is the field intensity. Thus, in spatial regions where the longitudinal component of the phase gradient becomes negative one should expect an energy backflow. One of such areas is the region surrounding the vortex phase singularity point, in which optical current streamlines circulate [Berry2010].

Consider the Poynting vector dynamics in a solid microsphere as presented by the vector maps in the left column of Fig. 1. From Figs. 1(a) and (b) it is clear, that generally, three vortices denoted as $B_1$-A-$B_2$ with counter-circulation energy fall into the considered region. In the WGM resonance (Fig. 1(a)), two oppositely directed vortices A and $B_2$ form a phase field *saddle*, which is located right at the microparticle rim. Therefore, at the particle edge the longitudinal energy flow is practically absent at WGM resonance: $S_z(y = 0) \to 0$. As a result, near the internal shadow surface of the particle the axial Poynting flux is directed outward, i.e. along the optical radiation, while on the outer edge side the energy flux is reversed and directed backward.

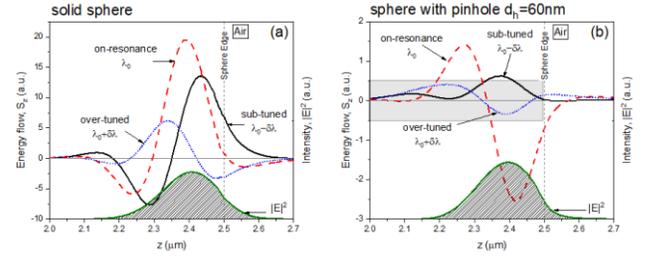

Fig. 2. The behavior of the longitudinal Pointing vector component $S_z$ near the rim of a (a) solid and (b) perforated sphere when tuned to and slightly detuned from the WGM resonance. Resonant optical intensity $|\mathbf{E}|^2$ profile is shown by green shaded line.

This conclusion is illustrated in Fig. 2(a), where the distribution of the longitudinal Poynting component $S_z$ near the edge of a solid microsphere is plotted along its diameter ($y = 0$). As can be seen, on the mode resonance $\lambda = \lambda_0$ (red dashed line) outside the particle $S_z < 0$; near the internal sphere edge the Pointing component is positive, $S_z > 0$, and on sphere boundary at $z = 2.5$ μm one has $S_z \approx 0$. At a small spectral detuning $\delta\lambda$ from the resonance (within the resonance bandwidth) the energy flux at the microparticle edge becomes nonzero and can change its direction. In the case of "blue" shift, $\lambda = \lambda_0 - \delta\lambda$, the Poynting flux is co-directed with the wave vector $k_z$, while at "red" shifted wavelength, $\lambda = \lambda_0 + \delta\lambda$, the energy flux is pulled into the particle. In the vector distributions shown in Figs. 1(a) and (b), the situations considered above are expressed by the longitudinal displacement (deepening) of the vortex triplet $B_1$-A-$B_2$ inside the particle at increasing optical wavelength $\lambda$. As a result, when the optical resonance is over-tuned, $\lambda > \lambda_0$, the center of the rightmost vortex $B_2$ penetrates the particle, and because of the left-handed circulation creates a backward energy flux moving into the microsphere interior.

For a microsphere with an open-end hole, the corresponding Pointing flux distributions are shown in Figs. 1(c) and (d), with the position of the hole conventionally shown by a shaded rectangle. First of all, it can be noted that from the viewpoint of Poynting vector field topology, the nano-hole inverts the optical vortex triplet from B-A-B to A-B-A due to the appearance of the WGM field reflection from the particle-air interface. Worthwhile noting, the pull-in component of the energy backflow near the boundary of a solid particle is zero, and it appears only in the presence of a hole.

Additionally, the character of the $A_1$-B-$A_2$ vortex triplet migration upon resonance detuning is different from that of a solid sphere. Indeed, at optical resonance, $\lambda = \lambda_0$, the phase field saddle between the B-$A_2$ vortices is located near the outer surface of dielectric particle in the hole region. This leads to the appearance of a weak but inversely directed Poynting flux at the hole opening, which then increases and reaches a maximum inside the hole at a deepening of $\Delta z$ = 80 nm (~$\lambda_0$/7) and then changes its sign at $\Delta z$ = 150 nm (see Fig. 2(b)). A small detuning from the resonance (Fig. 1(d)) changes both the mutual arrangement of the vortices and their spatial sizes. In this case, the saddle of optical current streamlines between the vortices B-$A_2$ shifts inside the particle as the wavelength detuning becomes "red". This spatial shift of the singularity points causes different behavior of $S_z$ component near the edge of the microsphere when the incident wavelength is changed. However, generally, in the off-resonance condition, the amplitude of the longitudinal Poynting flux at the hole opening is much smaller than that at WGM resonance.

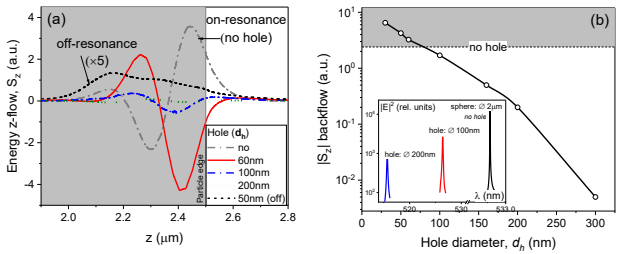

Fig. 3. (a) Longitudinal Poynting component $S_z$ inside a hole in the shadow microsphere side in the dependence of hole diameter $d_h$. (b) Energy backflow amplitude $|S_z|$ at WGM resonance versus hole diameter. Inset: $TE_{19,1}$ resonance spectral contour shift with hole width.

Fig. 3(a) shows the longitudinal Poynting component $S_z$ calculated near hole opening as a function of its diameter $d_h$. Here, $S_z(z)$ profile for a solid microparticle (without a hole) is plotted also at $TE_{19,1}$ - mode resonance (gray dashed curve), as well as for a particle with a 60 nm (~0.1$\lambda$) air hole but not in resonance, as in Fig. 1(a).

Noteworthy, when incident radiation is not in resonance with particle WGM (dashed line in Fig. 3(a), ×5 magnification), the longitudinal component of the Poynting vector is always positive throughout the hole, $S_z > 0$, and no reverse energy current is organized within the hole zone. However, on-resonance, the small positive $S_z$ flux in air sharply changes the sign and becomes opposite even near the hole exit plane in air (at $z$ = 2.5 μm). Meanwhile, for a 60 nm air hole, the ratio of $|S_z|$ amplitudes in the regions of negative (at $z$ = 2.4μm) and positive (at $z$ = 2.26μm) energy current amounts to about two. This is close in magnitude to the corresponding ratio in the case with the focusing of a second-order cylindrical vector beam by a Mikaelian gradient lens with a fabricated air crater [Stafeev2020]. However, the optical scheme considered in this paper is significantly simpler to implement and does not require specially structured radiation and specific complex structure of the refractive index (by hyperbolic cosine as in [Stafeev2020]) of a spherical particle.

The peak $S_z$ - values in the energy backflow region are summarized in Fig. 3(b) for different hole diameters. It should be noted that the fabrication of a hole in a solid dielectric particle changes the resonance conditions for the eigenmodes by shifting the resonant wavelength into the blue region. The inset to Fig. 3(b) shows the spectral contours of $TE_{19,1}$ WGM as the hole diameter changes. Qualitatively, this blue shift can be explained by a decrease in the effective refractive index $n_{eff}$ of a perforated microsphere: $n_{eff} \approx n(1 - V_h/V_0)$, where $V_0, V_h$ are the volumes of sphere and hole, respectively. Since the resonance Mie parameter $\rho_0 = 2\pi R/\lambda_0$ is inversely proportional to the refractive index of the particle [Weinstein] as $\rho_0 \approx (l^2 - 1)/n$, where $l$ is the mode number of the eigenmode (in this case $l$ = 19), then also the resonance wavelength $\lambda_0$ will be proportional to $n$ ($\lambda_0 \propto n$). Thus, considering the decrease in the effective refractive index $n_{eff}$ of the perforated particle one obtains that the change in the resonance WGM wavelength $\delta\lambda_0$ is proportional to the relative volume of the hole according to the relation: $\delta\lambda_0 \propto -V_h/V_0 \propto -(d_h/R)^2$.

Recalling Fig. 3(b) we note that the amplitude of the Pointing backflow continuously drops with increasing hole diameter. The functional fitting of this dependence results in an exponential law $|S_z| \propto \exp\{-ad_h\}$ with a decrement a of $a \approx 0.021$ nm, which correlates with a decrease in the WGM intensity in the hole region in the similar way $I_{max} \propto \exp\{-bd_h\}$, where by the calculation we have a slightly smaller, but close in magnitude decrement, $b \approx 0.019$ nm. This is clear, since the longitudinal component of the Poynting vector is proportional to the field intensity, $S_z \propto I$, and the intensity drop, in turn, is associated with the broadening of the spectral contour of the mode resonance and a decrease in its quality factor as $d_h$ increases. Therefore, tuning to a lower in quality WGM causes a decrease in the backflow amplitude in the area of hole outlet. Actually, according to our calculations, for $TE_{18,3}$-mode of a microsphere with the same 60 nm hole the amplitude of the reverse energy flux $|S_z|$ is almost 10 times lower than in the case shown in Fig. 3(a). Besides, for a 100 nm (~$\lambda$/5) hole this ratio is already more than 20 times. Thus, the highest values of the Poynting backflow component $|S_z|$ are realized for the highest-order eigen resonances of the particle.

In conclusion, a perforated mesoscale dielectric sphere, as well as a solid one, supports the electromagnetic high-quality eigenmodes – the whispering gallery modes. The influence of a hole results in a marked "blue" shift of WGM, lowering its quality and intensity as the hole diameter increases. By means of the numerically simulations, we show that at a WGM resonance, a strong backward flux of optical energy arises inside the hole directed towards the incident optical wave. The origin of this backflow is attributed to the multiple vortex singularities of optical wave phase arising near the particle rim at resonance. The intensity of resonant backflow decreases both as the hole diameter increases and the quality factor of the excited WGM drops, but remains multiple higher of the value realized in the nonresonant condition. By a small detuning from the resonance makes it possible to control the position of vortices near the nanohole, which in turn gains the control over the ratio of forward and backward optical energy fluxes in a mesoscale particle.

**Funding.** Ministry of Science and Higher Education of the Russian Federation (IAO SB RAS);

**Acknowledgments.** I.M. and O.M. author proposed the original idea, edit a paper. Y.G. wrote a draft and made simulations. Y.G. thanks the Ministry of Science and Higher Education of Russia for funding for this work. I.M. and O.M. thanks Tomsk Polytechnic University Competitiveness Enhancement Program for supporting this work.

**Disclosures.** The authors declare no conflicts of interest.

**Data availability**. No data were generated or analyzed in the presented research.

## References

1. A. Boivin, J. Dow, and E. Wolf, "Energy Flow in the Neighborhood of the Focus of a Coherent Beam*," J. Opt. Soc. Am. 57, 1171-1175 (1967).
2. H.F. Schouten, T.D. Visser, D. Lenstra, "Optical vortices near sub-wavelength structures," J. Opt. B: Quant. and Semic. Opt. **6**(5), S404-S409 (2004).
3. H. Li, C. Wang, M. Tang, and X. Li, "Controlled negative energy flow in the focus of a radial polarized optical beam," Opt. Express 28, 18607-18615 (2020)
4. S.S. Stafeev, E.S. Kozlova, A.G. Nalimov, V.V. Kotlyar, "Tight focusing of a cylindrical vector beam by a hyperbolic secant gradient index lens," Opt. Lett. **45**, 1687-1690 (2020).
5. H.Wang, J. Hao, B. Zhang, C. Han, C. Zhao, Z. Shen, J. Xu, and J. Ding, "Donut-like photonic nanojet with reverse energy flow," Chinese Opt. Lett. **19**(10), 102602 (2021).
6. B. S. Luk'yanchuk, R. Paniagua-Domínguez, I. V. Minin, O. V. Minin, and Z. Wang, "Refractive index less than two: photonic nanojets yesterday, today and tomorrow [Invited]," Opt. Mater. Express **7**, 1820–1847 (2017).
7. L. Chen, Y. Zhou, M. Wu, and M. Hong, "Remote-mode microsphere nano-imaging: new boundaries for optical microscopes," Opto-Electron. Adv. **1**, 170001 (2018).
8. M. X. Wu, B. J. Huang, R. Chen, Y. Yang, J. F. Wu, R. Ji, X. D. Chen, and M. H. Hong, "Modulation of photonic jets generated by microspheres decorated with concentric rings," Opt. Express **23**, 20096–20103 (2015).
9. Y. Zhou, H. Gao, J. Teng, X. Luo, and M. Hong, "Orbital angular momentum generation via a spiral phase microsphere," Opt. Lett. **43**, 34–37 (2018).
10. Y. Zhou and M. Hong, "Formation of polarization-dependent optical vortex beams via an engineered microsphere," Opt. Express **29**, 11121–11131 (2021).
11. Y. Cao, Z. Liu, O. V. Minin and I. V. Minin, "Deep subwavelength-scale light focusing and confinement in nanohole-structured mesoscale dielectric spheres," Nanomaterials **9**(2), 186 (2019).
12. I. V. Minin, O. V. Minin, Y. Cao, Z. Liu, Yu. E. Geints and A. Karabchevsky. "Optical vacuum cleaner by optomechanical manipulation of nanoparticles using nanostructured mesoscale dielectric cuboid." Sci. Rep. **9**:12748 (2019).
13. O. V. Minin, I. V. Minin, Y. Cao, "Optical magnet for nanoparticles manipulations based on optical vacuum cleaner concept," Proc. SPIE 11845, 118451G (2021).
14. A.Ya. Bekshaev, M.S. Soskin "Transverse energy flows in vectorial fields of paraxial beams with singularities," Opt. Commun. **271**, 332-348 (2007).
15. M.V. Berry "Optical currents," J. Opt. A: Pure Appl. Opt. **11,** 094001 (2009).
16. M.V. Berry, "Quantum backflow, negative kinetic energy, and optical retro-propagation," J. Phys. A: Math. Theor. **43**, 415302 (2010)